\documentclass[conference]{IEEEtran} 

\usepackage{etex}

\def\papertitle{An Early-Stopping Mechanism for DSCF Decoding of Polar Codes}

\usepackage{ifpdf}
\ifpdf
\usepackage[draft,pdfborder={0 0 0}]{hyperref}

\fi

\usepackage{cite}
\usepackage{graphicx}
\usepackage{epstopdf}

\usepackage{amssymb}
\usepackage[cmex10]{amsmath}
\usepackage[subnum]{cases}
\usepackage{nicefrac}
\usepackage{bm} 

\usepackage{algorithm}
\usepackage{algorithmicx}
\usepackage{algpseudocode}

\usepackage{glossaries}

\usepackage{tabularx}
\usepackage{multirow}
\usepackage{dcolumn}
\usepackage{booktabs}

\usepackage{tikz,pgf}
\usepackage{pgfplots}
\usetikzlibrary{plotmarks,spy}

\usepackage{balance}
\usepackage{caption}

\title{\papertitle}

\author{
  \IEEEauthorblockN{Ilshat Sagitov and Pascal Giard}
  \IEEEauthorblockA{
		LaCIME, Electrical Engineering Department,\\\'Ecole de technologie sup\'erieure (\'ETS), Montr\'eal, Canada\\
    Email: ilshat.sagitov.1@ens.etsmtl.ca, pascal.giard@etsmtl.ca
  }
}

\newcommand{\plotfigureheight}{0.75}

\makeglossaries

\begin{document}

\newacronym{awgn}{AWGN}{additive white Gaussian noise}
\newacronym{snr}{SNR}{signal-to-noise ratio}
\newacronym{fer}{FER}{frame-error rate}

\newacronym[plural=SNRs]{snr}{SNR}{signal-to-noise ratio}
\newacronym{sc}{SC}{successive-cancellation}
\newacronym{scl}{SCL}{successive-cancellation list}
\newacronym{scf}{SCF}{successive-cancellation flip}
\newacronym{dscf}{DSCF}{dynamic successive-cancellation flip}

\newacronym{crc}{CRC}{cyclic-redundancy check}
\newacronym{llr}{LLR}{log-likelihood ratio}

\maketitle

\begin{abstract}
Polar codes can be decoded with the low-complexity \gls{scf} algorithm. To improve error-correction performance, the \gls{dscf} variant was proposed, where the resulting error-correction performance is similar to that of the \glsentrylong{scl} algorithm with low to moderate list sizes. Regardless of the variant, the \gls{scf} algorithm exhibits a variable execution time with a high (worst-case) latency. In this work, we propose an early-stopping metric used to detect codewords that are likely undecodable such that the decoder can be stopped at earlier stages for those codewords. We then propose a modified version of the \gls{dscf} algorithm that integrates our early-stopping metric that exploits the specific properties of \gls{dscf}. Compared to the original \gls{dscf} algorithm, in the region of interest for wireless communications, simulation results show that our proposed modifications can lead to reductions of 22\% to the average execution time and of 45\% to the execution-time variance at the cost of a minor error-correction loss of approximately 0.05\,dB.
\end{abstract}

\glsresetall

\section{Introduction}
\label{sect:intro}
Polar codes, introduced in~\cite{arik_polariz}, are a type of linear error-correction codes which can achieve the channel capacity for practically relevant channels. The original \gls{sc} decoding algorithm has low complexity, but its error-correction performance is lacking for many practical applications. To address this, the \gls{scl} decoding algorithm was proposed. It provides great error-correction capability to the extent that polar codes were selected to protect the control channel in 3GPP's next-generation mobile-communication standard (5G), where the \gls{scl} algorithm serves as the error-correction performance baseline~\cite{3GPP_5G_Coding}. The great error-correction performance of \gls{scl} comes at the cost of high hardware implementation complexity and low energy efficiency~\cite{scl_5g}. 

As an alternative, the \gls{scf} decoding algorithm was proposed in~\cite{scf_intro}. The \gls{scf} algorithm can be summarized as the application of multiple \gls{sc} decoding trials, where each trial, beyond the first, flips the value of one intermediate decision. Which value---or bit---to flip is determined from a list that is meant to keep track of the least-reliable decisions, and the latency of an \gls{scf} decoder is a function of the maximum number trials. The \gls{scf} decoding algorithm leads to an improved error-correction performance compared to an \gls{sc} decoder, but still falls short of that of an \gls{scl} decoder with a moderate list size. However, compared to an \gls{scl} decoder, an \gls{scf} decoder is more efficient both in terms of computing resources and energy requirements~\cite{Giard_JETCAS_2017}.

The \gls{dscf} decoding algorithm was proposed in~\cite{dyn_scf}, where the authors introduced two major modifications to the original \gls{scf} algorithm with the goal of improving the error-correction performance. Firstly, the \gls{dscf} algorithm takes a new, more accurate, approach on the metric computations used to establish the list of the least-reliable decisions. Secondly, that algorithm allows for multiple bit flips per trial. With these modifications, the error-correction performance becomes close to that of \gls{scl} decoder with small to moderate list sizes~\cite{simp_dscf}.

Regardless of these improvements, both \gls{scf} and \gls{dscf} decoders have a variable execution time by nature, and the variance on that execution time can be significant. This poses a challenge in the realization of receivers, where fixed-time algorithms are preferred.
A high variance on the execution time of the decoder leads to large buffers. Alternatively, the maximum number of trials can be reduced, dynamically or not, at the cost of a reduced error-correction performance. 
\subsubsection*{Contributions}
In this work, we present modifications to the original \gls{dscf} decoding algorithm to integrate an early-stopping mechanism that attempts to distinguish undecodable codewords from decodable ones based on a new early-stopping metric combined with a pre-calculated threshold. Based on that metric, a codeword may be classified as likely undecodable, in which case the decoder attempts a limited number of trials, much lower than the maximum number of trials. After those trials, if decoding is not successful, the decoder stops. The behavior of the decoder for the other codewords remains unaffected. As a result of our modifications to the \gls{dscf} algorithm, the average execution time and its variance are reduced at a cost of a minor error-correction performance loss. 

\subsubsection*{Outline} The remainder of this paper is organized as follows. Section~\ref{sect:backgr} provides a theoretical overview on polar codes, their construction and the \gls{sc} decoding algorithm. In addition to that, the \gls{scf} and \gls{dscf} decoding algorithms are briefly presented. In Section~\ref{sect:alg_modif}, the proposed modifications to the \gls{dscf} algorithm are explained, and the new metric along with its calculations used for early stopping is presented. In Section~\ref{sect:obt_thr}, the methodology for obtaining the metric distribution over the required number of trials and defining thresholds is explained. In Section~\ref{sect:results}, the proposed modified \gls{dscf} algorithm is compared against the original one in terms of execution time and error-correction performance. Section~\ref{sect:conclusion} concludes this work and proposes potential future works. 

\section{Background}
\label{sect:backgr}
\subsection{Construction of Polar Codes}
Polar codes are based on the concept of channel polarization. As the code length tends to infinity, bit locations either become completely reliable or completely unreliable. To construct a $\mathcal{P}\left(N,\,k\right)$ polar code, where $N$ is the code length and $k$ the number of information bits, the $\left(N-k\right)$ least-reliable bits are set to predefined values, typically all zeros. These bits are called the frozen bits. The sets of information- and frozen-bit locations are often denoted as $\mathcal{A}$ and $\mathcal{A}^C$, respectively. The encoding is the linear transformation such that $\bm{x}=\bm{u}\times F^{\otimes n}$, where $\bm{x}$ is the polar-encoded row vector, $\bm{u}$ is a row vector of length $N$ that contains the $k$ information bits in their predefined locations as well as the frozen bit values, $n=\log_2 N$ and $F^{ \otimes n}$ is the $n^{th}$ Kronecker product $\left( \otimes \right)$ of the binary polar-code kernel $F=\scriptstyle{\begin{bmatrix} 1 & 0 \\ 1 & 1 \end{bmatrix}}$. The bit-location reliabilities depend on the channel type and conditions. In this work, the  \gls{awgn} channel is considered and the polar-code construction method used is that of Tal and Vardy~\cite{tal_constr}. 

\subsection{Successive-Cancellation Decoding}
\begin{figure}[ht]
	\centering
	\begin{tikzpicture}[]

\node [draw, circle, line width=0.8pt, inner sep=0.1 cm, label=south:{$\hat{u}_0$}] (u0) at (0,0){};
\node [draw, circle, line width=0.8pt, inner sep=0.1 cm, label=south:{$\hat{u}_1$}] (u1) at (0.6,0){};
\node [draw, circle, line width=0.8pt, inner sep=0.1 cm, label=south:{$\hat{u}_2$}] (u2) at (1.2,0){};
\node [draw, circle, line width=0.8pt, inner sep=0.1 cm, label=south:{$\hat{u}_3$}, fill=black] (u3) at (1.8,0){};
\node [draw, circle, line width=0.8pt, inner sep=0.1 cm, label=south:{$\hat{u}_4$}] (u4) at (2.4,0){};
\node [draw, circle, line width=0.8pt, inner sep=0.1 cm, label=south:{$\hat{u}_5$}, fill=black] (u5) at (3.0,0){};
\node [draw, circle, line width=0.8pt, inner sep=0.1 cm, label=south:{$\hat{u}_6$}, fill=black] (u6) at (3.6,0){};
\node [draw, circle, line width=0.8pt, inner sep=0.1 cm, label=south:{$\hat{u}_7$}, fill=black] (u7) at (4.2,0){};

\node [draw, circle, line width=0.8pt, inner sep=0.1 cm, fill=gray!50!white] (n10) at (0.3,0.5){};
\node [draw, circle, line width=0.8pt, inner sep=0.1 cm, fill=gray!50!white] (n11) at (1.5,0.5){};
\node [draw, circle, line width=0.8pt, inner sep=0.1 cm, fill=gray!50!white] (n12) at (2.7,0.5){};
\node [draw, circle, line width=0.8pt, inner sep=0.1 cm, fill=gray!50!white] (n13) at (3.9,0.5){};
\node [draw, circle, line width=0.8pt, inner sep=0.1 cm, fill=gray!50!white] (n20) at (0.9,1.0){};
\node [draw, circle, color=green,  line width=0.8pt, inner sep=0.05 cm, fill=gray!50!white] (n21) at (3.3,1.0){\textcolor{black}{\footnotesize{$v$}}};
\node [draw, circle, line width=0.8pt, inner sep=0.1 cm, fill=gray!50!white] (n30) at (2.1,1.5){};
\draw (u0) -- node{}(n10);
\draw (u1) -- node{}(n10);
\draw (u2) -- node{}(n11);
\draw (u3) -- node{}(n11);
\draw (u4) -- node{}(n12);
\draw (u5) -- node{}(n12);
\draw (u6) -- node{}(n13);
\draw (u7) -- node{}(n13);
\draw [line width=0.8pt] (n10) -- node{}(n20);
\draw [line width=0.8pt] (n11) -- node{}(n20);
\draw [line width=0.8pt] (n12) -- node{}(n21);
\draw [line width=0.8pt] (n13) -- node{}(n21);
\draw [line width=1.2pt] (n20) -- node{}(n30);
\draw [line width=1.2pt] (n21) -- node{}(n30);
\draw [-latex] (2.3,1.56) -- node[anchor=south] {\footnotesize{$\alpha_v$}}(3.2,1.2);

\draw [-latex] (3.1,0.95) -- node{} (2.75,0.68);
\draw [-latex] (2.9,0.55) -- node{} (3.2,0.79);
\draw [-latex] (3.5,0.95) -- node{} (3.85,0.68);
\draw [-latex] (3.7,0.55) -- node{} (3.4,0.79);

\node [] (al) at (2.8,0.93){\footnotesize{$\alpha_l$}};
\node [] (ar) at (3.8,0.93){\footnotesize{$\alpha_r$}};
\node [] (bl) at (3.1,0.45){\footnotesize{$\beta_l$}};
\node [] (br) at (3.6,0.45){\footnotesize{$\beta_r$}};

\end{tikzpicture}
	\caption{Tree-based representation of $\mathcal{P}\left( 8,\,4\right)$.}
	\label{fig:pol_tree}
\end{figure}
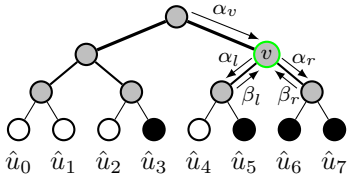

Binary trees are a natural representation of polar codes. Fig.\,\ref{fig:pol_tree} illustrates the tree-based representation of a $\mathcal{P}\left(8,\, 4\right)$ polar code. A polar code of length $N$ can be seen as a composition of two smaller constituent polar codes of length $N/2$.
The color of the leaves indicate whether a location is that of a frozen or information bit. Frozen bits are colored white and information bits are set to black.  The \gls{sc} decoding algorithm follows the depth-first traversing schedule of the tree, visiting the left child nodes first and then passing the calculated messages to the right child nodes. The messages passed to the child nodes are vectors of \glspl{llr}, while those passed to parents are vectors of bits, often called partial-sum bits. These messages are denoted as $\alpha$ and $\beta$, respectively. 

By taking the node indicated with $v$ in the decoding tree, the messages to its left $l$ and right $r$ children are calculated according to Eq.\,\eqref{eq:min_sum}. The calculations of $f$ and $g$ functions are provided in Eq.\,\eqref{eq:f_g_func} for arbitrary real values $x$ and $y$, and binary value $b$.

\begin{equation}
\begin{split}
\alpha_l[i] &= f\left(\alpha_v\left[i\right],\, \alpha_v\left[i+N_v/2\right] \right) \\
\alpha_r[i] &= g\left(\alpha_v\left[i\right],\, \alpha_v\left[i+N_v/2\right],\,\beta_l\left[i\right] \right)
\end{split}
\label{eq:min_sum}
\end{equation}

\begin{equation}
\begin{split}
&f\left(x,\,y\right) = \mathrm{sign}\left( x\right) \cdot \mathrm{sign} \left( y \right) \cdot \mathrm{min}\left( |x|,\,|y| \right) \\
&g\left(x,\,y,\,b\right) = \left( 1-2b\right) \cdot x + y 
\end{split}
\label{eq:f_g_func}
\end{equation}

The value $\beta_l$ in Eq.\,\eqref{eq:min_sum} is the estimated bit vector coming to the node $v$ from its left child. In the leaf nodes the estimated bits are calculated with a hard decision on the input \glspl{llr}. These \glspl{llr} are called the decision \glspl{llr} and are denoted as $\alpha_{dec}$. In the case of frozen bits, hard decisions are not needed, and those bits are directly set to the all-zero vector. For the intermediate nodes, the partial-sum bits are calculated as follows:

\begin{equation}
\beta_v\left[i\right] = 
\begin{cases}
\beta_l\left[i\right] \oplus \beta_r\left[i\right] ~~~ \mathrm{if} ~~ i < \frac{N_v}{2}\,,\\
\beta_r\left[i\right] \quad \quad \quad ~~~\, \mathrm{otherwise}\,.
\end{cases}
\label{eq:bit_esimates}
\end{equation}

\subsection{Successive-Cancellation Flip Decoding}
In~\cite{scf_intro}, at first, the authors designed an Oracle-assisted decoder and showed that the error-correction performance would significantly improve if the first erroneously-estimated bit could be detected and corrected before resuming \gls{sc} decoding. Then based of these results, they proposed the \gls{scf} decoding algorithm. In order to detect the decoding failure of the codeword, the information bits are concatenated with \gls{crc} bits of the length $r$ thus increasing the code rate of the polar code to $R=\left(k+r\right)/N$. The list of bit-flipping candidates $\mathcal{L}_{flip}$ is constructed, if its \gls{crc} check fails after the first \gls{sc} decoding pass, based on the least-reliable bit indices, i.e., indices with the smallest absolute $\alpha_{dec}$ values of the non-frozen bits after the first \gls{sc} trial. For each new \gls{sc} trial the next bit index of $\mathcal{L}_{flip}$ is chosen and when this bit is estimated, the opposite decision is made, i.e., the estimated bit is flipped. Decoding is considered successful if the \gls{crc} matches. In order to constrain the latency, a maximum number of trials $T$ is defined, where $T$ is an integer number such that $0\leq T < \left( k+r\right)$. Setting $T$ to 0 renders the \gls{scf} decoder equivalent to an \gls{sc} decoder. The list of bit-flipping candidates consists of $\mathcal{L}_{flip}=\{i_1,
\, i_2,\, \ldots i_T\}$, where $i$ is the index of the bit-flipping candidate. The metric associated to each non-frozen bit is defined as:
\begin{equation}
M_i=|\alpha_{dec}[i]|\,.
\label{eq:metr_scf}
\end{equation}

The resulting metrics are sorted in ascending order and the first $T$ metrics are chosen to form the metric list $M^T$. Then the list of the bit-flipping candidates $\mathcal{L}_{flip}$ associated to the metric list is constructed. The \gls{scf} decoder has $T$ additional trials beyond the initial \gls{sc} pass to decode the codeword while applying the next bit from $\mathcal{L}_{flip}$ each trial. If decoding is not successful after trial $T$, decoding is stopped and the codeword considered undecodable.

\subsection{Dynamic Successive-Cancellation Flip Decoding}
The \gls{dscf} decoder constructs the list of the bit-flipping candidates differently compared to the \gls{scf} algorithm~\cite{dyn_scf}. The metric calculated for each bit is the probability of the event that this bit is wrongly estimated and that every previous non-frozen bits are estimated correctly. In other words, the metric calculated for the current bit indicates the probability that this bit is the first channel-induced error which cannot be corrected by the \gls{sc} decoding algorithm~\cite{dyn_scf}. The metric can be calculated in both linear or logarithmic domain. Throughout this work, the logarithmic domain computations are considered. The metric calculation for each non-frozen bit after the initial \gls{sc} attempt is defined as:
\begin{equation}
M_i = |\alpha_{dec}[i]| + \frac{1}{c} \cdot  \sum_{\substack{j \leq i \\ j\in \mathcal{A}}} \ln \left( 1 + e^{\left( -c \cdot |\alpha_{dec}[j]|\right)}\right)\,,
\label{eq:metr_dscf}
\end{equation}
where $\ln(\cdot)$ denotes the natural logarithm, and $c$ is a constant optimized experimentally by way of simulation. The value $c$ will vary depending on the polar code parameters and channel conditions. 

In addition to the change in metric computations, the \gls{dscf} decoding algorithm allows for multiple bit flips per trial. Instead of a single bit index as an element of the list of bit-flipping candidates, now the set of bit indices is considered as one element. One set is denoted with $\epsilon$, $\epsilon = \{i_1,\,i_2,\,\ldots i_{\omega} \}$ and $\mathcal{L}_{flip}=\{\epsilon_1, \epsilon_2,\, \ldots\, \epsilon_{T}\}$, where $\omega$ is called the order of the decoder, i.e., the maximum number of bit flips within one candidate set. For calculation of metrics associated to sets, we refer the reader to \cite{dyn_scf} as they are not used in this paper. Indeed, in this work, $\omega=1$ is considered, and the decoding algorithm used could thus be denoted as \gls{dscf}-1. However, for the sake of simplicity, in the remainder of this paper, \gls{dscf} is also synonymous of the \gls{dscf}-1 algorithm unless stated otherwise.

\section{Proposed Algorithmic Modifications}
\label{sect:alg_modif}

\subsection{Introduction of the Early-Stopping Metric}
The core idea of our proposed method is an early-stopping mechanism that attempts to distinguish the undecodable codewords from the decodable ones. When a codeword that is likely undecodable is identified, our decoder uses a lower maximum number of trials. The early-stopping metric $\phi$ used to identify undecodable codewords is the variance of the elements of $M$ associated to the list of bit-flipping candidates. It is calculated as follows:
\begin{equation}
\phi = V_{M} = \frac{1}{T-1} \sum_{i=1}^{T} \left( m_i - m_{av}\right)^2\,,
\label{eq:var_metr_calc}
\end{equation}
where $m_i$ is the metric associated to the bit index from $\mathcal{L}\left[i \right]$, and $m_{av}$ is the mean of the elements of $M$ calculated as:
\begin{equation}
m_{av} = \frac{1}{T} \sum_{i=1}^{T}  m_i\,.
\label{eq:mean_metr_calc}
\end{equation}

The early-stopping metric $\phi$ is not to be confused with metrics associated with the bit-flipping candidates of the \gls{dscf} decoder. The metric $\phi$ of our proposed algorithm aims to distinguish the codewords that are likely undecodable from the decodable ones.

\subsection{\Gls{dscf} Decoder with Early Stopping}
The early-stopping metric $\phi$ explained in previous subsection can be integrated to the \gls{dscf} decoder and is calculated for the current codeword after obtaining the list $M$. Then the resulting metric is compared against the pre-calculated threshold metric $\phi_{thr}$. If $\phi > \phi_{thr}$, the maximum number of trials becomes the pre-defined reduced maximum number of trials $T=T_{red}$. Otherwise, the initial maximal number of trials $T$ is kept. The simulation set-up is created and its corresponding algorithm is summarized in Alg.\,\ref{alg:dscf_red_tr}.

\begin{algorithm}[H]
\caption{\gls{dscf} decoding algorithm with early stopping.}
\label{alg:dscf_red_tr}
\begin{algorithmic}[1]\small
\Procedure{Perform {\gls{dscf}}}{$T$,\, $\phi_{thr}$,\, $T_{red}$}
\State$\left(\hat{u}^N,\, \alpha_{dec}^N\right) \gets \mathrm{\gls{sc}}\left(\alpha_{ch}^N,\, \mathcal{A} \right)$
\If{$\gls{crc}\left(\hat{u}^N\right)=\mathrm{failure}$}
\State $\mathrm{Init}\left(\mathcal{L}_{flip}, \, M^T \right)$
\State $\phi \gets V_{M} \left( M^T \right)$
\If{$\phi > \phi_{thr}$}
\State $T\gets T_{red}$
\EndIf
\State $\left(\hat{u}^N,\, \alpha_{dec}^N,\, t \right) \gets \mathrm{\gls{scf}}\left(\alpha_{ch}^N,\, \mathcal{A},\, \mathcal{L}_{flip},\, T\right)$
\EndIf
\State \textbf{return} $\left[\hat{u}^N, \, t \right]$
\EndProcedure
\end{algorithmic}
\end{algorithm}

Simulations are for various \gls{snr} points. The output number of trials $t$ can be used to build statistics of the execution time averaged over the number of simulation cycles (simulated codewords). The average execution time $T_{av}$, in terms of average of number of trials, can be calculated as:
\begin{equation}
T_{av} = \frac{1}{S} \sum_{i=1}^{S}  t_i \,,
\label{eq:av_lat}
\end{equation}
where $S$ corresponds to the number of simulated codewords per \gls{snr} point.

The variance of the execution time $V_{T}$ can also be obtained from the output $t$ and is calculated as follows:
\begin{equation}
V_{T} = \frac{1}{S-1} \sum_{i=1}^{S} \left( t_i - T_{av}\right)^2 \,.
\label{eq:var_lat}
\end{equation}

Simulation results are provided in Section~\ref{sect:results}.

\section{Obtaining the Metric Threshold from the Early-Stopping Metric Distribution}
\label{sect:obt_thr}
The \gls{dscf} algorithm with an early stopping requires the threshold metric $\phi_{thr}$ and the reduced maximum number of trials $T_{red}$, values determined by way of simulation. At first, the distribution of the average $\phi$ is obtained as a function of the number of trials $t$ required by the original \gls{dscf} decoder. Then the corresponding thresholds can be defined based on the resulting distribution.

\subsection{Methodology}
The simulation setup created for obtaining the distribution of the average metric depending on $t$ uses random codewords encoded by a $\mathcal{P}\left(1024,\,512\right)$ polar code and a \gls{crc} of $r=16$ bits with polynomial $z^{16}+z^{15}+z^2+1$. Those polar-code length and rate, and that \gls{crc} length were chosen to ease comparison with other works as they are commonly found in the literature. The polar code is constructed for an approximate design \gls{snr} $\gamma_{des}$ of $2.365$\,dB. Binary phase-shift keying modulation is used over an \gls{awgn} channel. For each \gls{snr} point, the simulation is run for $10^7$ codewords. The maximum number of trials beyond the initial \gls{sc} decoding pass of the \gls{dscf} decoder is $T=10$. The value $c$ used in the calculation of the \gls{dscf} metrics associated to the bit-flipping candidates of Eq.\,\eqref{eq:metr_dscf} is $0.3$.

\subsection{Average Early-Stopping Metric Distribution}
Alg.\,\ref{algo:metr_distr_dscf} shows the algorithm used to obtain the average early-stopping metric distribution, where $s$ stands for the simulation index and $S$ corresponds to the total number of simulated codewords. The list of metrics $\Phi$ has length $T+2$ and includes the resulting early-stopping metrics obtained from the successful initial \gls{sc} trials $t = 0$, those corresponding to undecodable codewords, and those for each $1 \leq t \leq T$ corresponding to successfully-decoded codewords. It can be seen that the metric $\phi$ is calculated on line 8 and is accumulated into the list $\Phi$ at the appropriate position based on $t$, either on line 13 or 16 if decoding is successful or not, respectively. The list of counters $C$, which has the same length as $\Phi$, increments its element according to the output $t$ on line 17 in the case of successful decoding or on line 14 otherwise. At the end of the simulation, the elements of the list of accumulated metrics $\Phi$ are normalized by their corresponding counters to obtain the average value. This simulation is ran for each \gls{snr} point. 

\begin{algorithm}[t]
\caption{Obtaining the average metric distribution.}
\label{algo:metr_distr_dscf}
\begin{algorithmic}[1]\small
\Procedure{Get\_distribution}{$S$}
\State $\Phi^{T+2} \gets \{0,\,0,\,\ldots, 0\} $
\State $C^{T+2} \gets \{0,\,0,\,\ldots, 0\}$
\For{$s=1$;\, $s \leq S$;\, $s++$}
	\State $\left(\hat{u}_1^{N},\, \alpha_{dec}^{N}\right)\gets \mathrm{\gls{sc}}\left(\alpha_{ch}^{N},\, \mathcal{A} \right)$
	\State $t \gets 0$
	\State $\mathrm{Init}\left(\mathcal{L}_{flip}, \, M^{T} \right)$
	\State $\phi \gets V_{M}\left( M^{T} \right)$
	\If{$\gls{crc}\left(\hat{u}_1^{N}\right)=\mathrm{failure}$}
		\State $\left(\hat{u}^{N},\, \alpha_{dec}^{N},\, t \right) \gets \mathrm{\gls{scf}}\left(\alpha_{ch}^{N},\, \mathcal{A},\, \mathcal{L}_{flip},\, T\right)$
	\EndIf
	\If{$\left(u_1^{N} \neq \hat{u}_1^{N}\right)$}
		\State $\Phi\left(T+1\right) \gets \Phi\left(T+1\right) + \phi$
		\State $C\left(T+1\right) \gets C\left(T+1\right) + 1$
	\Else 
		\State $\Phi\left(t\right) \gets \Phi\left(t\right) + \phi$
		\State $C\left(t\right) \gets C\left(t\right) + 1$
	\EndIf
\EndFor
\For{$k=0$;\, $k \leq \left(T+1\right)$;\, $k++$}
\State $\Phi\left(k \right) \gets \frac{\Phi\left(k\right)}{C\left(k\right)}$
\EndFor
\State \textbf{return} $\Phi$
\EndProcedure
\end{algorithmic}
\end{algorithm}

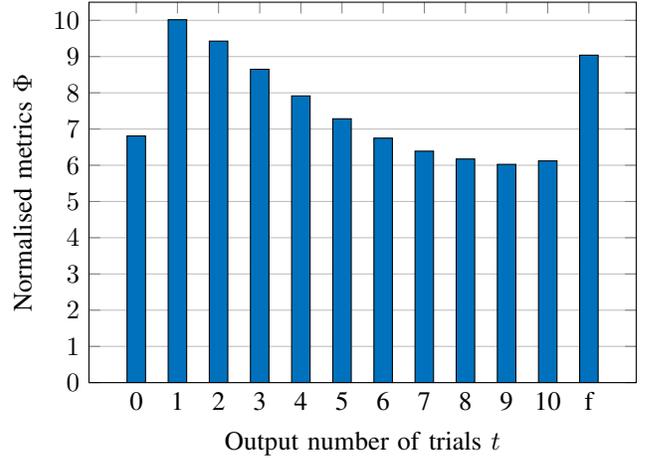
\begin{figure}[ht]
	\definecolor{mycolor1}{rgb}{0.00000,0.44700,0.74100}%
\begin{tikzpicture}

  \begin{axis}[%
    width=\columnwidth,
    height=\plotfigureheight\columnwidth,
    bar shift auto,
    xmin=-0.15, xmax=13.15,
    xtick={1,2,3,4,5,6,7,8,9,10,11,12},
    xticklabels={{0},{1},{2},{3},{4},{5},{6},{7},{8},{9},{10},{f}},
    xlabel={Output number of trials $t$},
    ymin=0, ymax=10.5,
    ytick={0,1,2,...,10},
    ylabel style={yshift=-1.0em},
    ylabel={$\text{Normalised metrics }\Phi$},
    ymajorgrids,
    ]
    \addplot[ybar, bar width=7.0, fill=mycolor1, draw=black, area legend] table[row sep=crcr] {%
      1	6.80853235150886\\
      2	10.0169989303445\\
      3	9.42680722266555\\
      4	8.64919234106904\\
      5	7.9148106409248\\
      6	7.28315629855536\\
      7	6.749035035249\\
      8	6.38904908741359\\
      9	6.17292166068686\\
      10	6.02504634992377\\
      11	6.12024996708021\\
      12	9.03870141945518\\
    };

  \end{axis}

\end{tikzpicture}%
	\caption{Distribution of early-stopping metrics depending on the number of trials required by the original \gls{dscf} decoder for a $\mathcal{P}\left(1024,\,512\right)$ polar code, a \gls{crc} of $r=16$ bits, and a \gls{awgn} channel with \gls{snr} $\gamma=2.25$\,dB.}
	\label{fig:metr_distr_dscf_2_25}
\end{figure}

The bar diagram of the resulting average early-stopping metric distribution is presented in Fig.\,\ref{fig:metr_distr_dscf_2_25} for $\gamma=2.25$\,dB. From this figure, it can be seen that the average early-stopping metrics are distributed exponentially within the range $1\leq t \leq 10$. The average early-stopping metric corresponding to undecodable codewords is clearly higher than those of decodable codewords with the highest values of $t$ and is close to that of $t=2$. 

\subsection{Defining the Threshold Metrics and the Reduced Number of Maximum Trials}
According to the distribution of the average early-stopping metrics shown in Fig.\,\ref{fig:metr_distr_dscf_2_25}, the threshold metric $\phi_{thr}$ gets the value of the average metric obtained for the undecodable codewords (denoted as "f" in figure). A threshold metric $\phi_{thr}$ is defined for each \gls{snr} point of interest. The relative position of the normalized metric for the undecodable codewords has been observed to be independent from the channel \gls{snr}. As a consequence, a single value for the reduced maximum number of trials $T_{red}$. In the following, simulation results will be presented and discussed for $T_{red}\in\{2,\,3,\,4\}$.

\section{Simulation Results}
\label{sect:results}
In this section, we compare the performance of our proposed modified \gls{dscf} decoding algorithm against that of the original algorithm. The comparison is made in terms of average execution time, execution-time variance, and error-correction performance.

\subsection{Methodology}
The results are obtained using the same parameters as those described in Section~\ref{sect:obt_thr} except that the simulations are run for $10^5$ blocks or until the targeted minimum number of blocks in error is reached. The minimum number of errors is $10^3$ for an \gls{snr} range of $\{1.0,\,1.25,\,\ldots\,2.25\}$\,dB. Then the minimum number of errors is $500$ and $300$ for \gls{snr} points of $2.5$\,dB and $2.75$\,dB accordingly. Several reduced numbers of maximum trials $T_{red}\in \{2,\,3,\,4\}$ are simulated for the \gls{dscf} decoder with the proposed early-stopping mechanism. The maximum number of trials beyond the initial \gls{sc} decoding pass of the \gls{dscf} decoder is $T=10$. The value $c$ used in the calculation of the metrics associated to the bit-flipping candidates of Eq.\,\eqref{eq:metr_dscf} is $0.3$.

\subsection{Average Execution Time and Variance}
\begin{figure}[t]
	\tikzset{new spy style/.style={spy scope={%
 magnification=5,
 size=1.25cm, 
 connect spies,
 every spy on node/.style={
   rectangle,
   draw,
   },
 every spy in node/.style={
   fill=white,
   draw,
   rectangle
   }
  }
 }
} 

\begin{tikzpicture}[new spy style]

  \begin{semilogyaxis}[%
    width=\columnwidth,
    height=\plotfigureheight\columnwidth,
    xmin=1, xmax=2.75,
    xlabel={$\gamma\left[ \mathrm{dB} \right]$},
    xtick={1,1.25,...,3.0},
    ymin=1e-2, ymax=10,
    ylabel style={yshift=-0.6em},
    ylabel={Average number of trials $T_{av}$},
    xmajorgrids, ymajorgrids,
    yminorticks, yminorgrids,
    legend style={legend cell align=left, align=left},
    mark size=3.0pt, mark options=solid,
    ]
    \addplot [color=black, line width=1.0pt, mark=o]
    table[row sep=crcr]{%
      1	7.80133\\
      1.25	5.71208\\
      1.5	3.41852\\
      1.75	1.58637\\
      1.875	1.00231\\
      2	0.59612\\
      2.125	0.33189\\
      2.25	0.1837200783\\
      2.5	0.0533\\
      2.75	0.0148545886\\
    };
    \addlegendentry{Original}
    
    \addplot [color=orange, line width=1.0pt, mark=star]
    table[row sep=crcr]{%
      1	4.29586\\
      1.25	3.19339\\
      1.5	1.97954\\
      1.75	0.98076\\
      1.875	0.64043\\
      2	0.39577\\
      2.125	0.2309\\
      2.25	0.13372\\
      2.5	0.0432362468\\
      2.75	0.0126391559\\   
    };
    \addlegendentry{$T_{red} = 2$}

    \addplot [color=red, line width=1.0pt, mark=diamond]
    table[row sep=crcr]{%
      1	4.74892\\
      1.25	3.5322\\
      1.5	2.18679\\
      1.75	1.0764\\
      1.875	0.7016\\
      2	0.43131000000\\
      2.125	0.24994000000\\
      2.25	0.14315971600\\
      2.5	0.04493183650\\
      2.75	0.0132033918\\     
    };
    \addlegendentry{$T_{red} = 3$}
    
    \addplot [color=blue, line width=1.0pt, mark=triangle]
    table[row sep=crcr]{%
      1	5.1953\\
      1.25	3.85969\\
      1.5	2.38016\\
      1.75	1.16128\\
      1.875	0.75358\\
      2	0.46063\\
      2.125	0.2648\\
      2.25	0.1511850686\\
      2.5	0.0467060133\\
      2.75	0.0135647475\\  
    };
    \addlegendentry{$T_{red} = 4$}
    \coordinate (lookhere) at (axis cs:2.25,1.5e-1);
    \spy[size=1.75cm,magnification=2.5] on (lookhere) in node at (1.1,1.1);
  \end{semilogyaxis}

\end{tikzpicture}%
	\caption{Average number of trials, beyond the first \gls{sc} decoding pass, for various \gls{snr} points for a \gls{dscf} decoder for $\mathcal{P}\left(1024,\,512\right)$ and $r=16$ with and without the proposed early-stopping mechanism.}
	\label{fig:av_exec}
\end{figure}
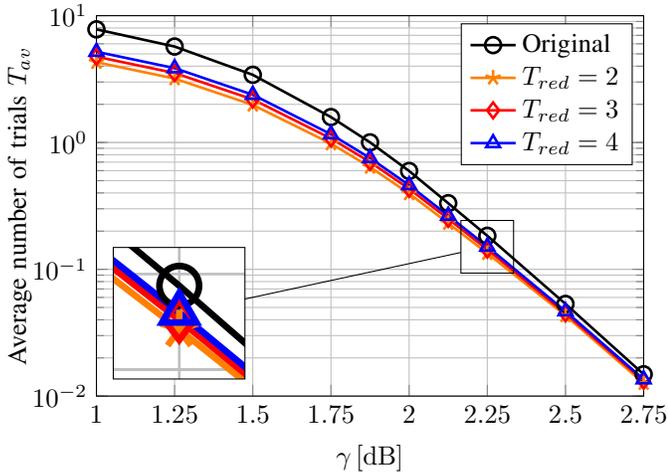

\begin{figure}[t]
	\tikzset{new spy style/.style={spy scope={%
 magnification=5,
 size=1.25cm, 
 connect spies,
 every spy on node/.style={
   rectangle,
   draw,
   },
 every spy in node/.style={
   fill=white,
   draw,
   rectangle
   }
  }
 }
} 

\begin{tikzpicture}[new spy style]

  \begin{semilogyaxis}[%
    width=\columnwidth,
    height=\plotfigureheight\columnwidth,
    xmin=1, xmax=2.75,
    xlabel={$\gamma\left[ \mathrm{dB} \right]$},
    xtick={1,1.25,...,3.0},
    ymin=1e-2, ymax=1e2,
    ylabel style={yshift=-0.6em},
    ylabel={Variance of the number of trials $V_{T}$},
    xmajorgrids, yminorticks,
    ymajorgrids, yminorgrids,
    legend style={legend cell align=left, align=left},
    mark size=3.0pt, mark options=solid
    ]
    \addplot [color=black, line width=1.0pt, mark=o]
    table[row sep=crcr]{%
      1	15.43527458472747860\\
      1.25	21.71919926597056971\\
      1.5	19.27213373133291441\\
      1.75	10.71862741026003185\\
      1.875	6.84603312434226030\\
      2	4.02668121281084268\\
      2.125	2.11422017020400022\\
      2.25	1.06935520381038773\\
      2.5	0.25092414474605418\\
      2.75	0.05468523296586780\\      
    };
    \addlegendentry{Original}
    
    \addplot [color=orange, line width=1.0pt, mark=star]
    table[row sep=crcr]{%
      1	16.05142737628040805\\
      1.25	14.36209392892349612\\
      1.5	10.24948388482479089\\
      1.75	5.33904321442837215\\
      1.875	3.34969291293238758\\
      2	1.982215930159821\\
      2.125	1.05051569515478671\\
      2.25	0.54430440504322242\\
      2.5	0.14133207011579749\\
      2.75	0.03069624863518522\\         
    };
    \addlegendentry{$T_{red} = 2$}

    \addplot [color=red, line width=1.0pt, mark=diamond]
    table[row sep=crcr]{%
      1	14.21888102281579158\\
      1.25	13.7773996460083783\\
      1.5	10.4222637186342233\\
      1.75	5.6205561895682\\
      1.875	3.57345317453473088\\
      2	2.1305229892312\\
      2.125	1.13656136361203042\\
      2.25	0.58837497868758704\\
      2.5	0.14658776368924006\\
      2.75	0.03336533477329826\\    
    };
    \addlegendentry{$T_{red} = 3$}
    
    \addplot [color=blue, line width=1.0pt, mark=triangle]
    table[row sep=crcr]{%
      1	12.90462695626447953\\
      1.25	13.64905959461223972\\
      1.5	10.89274730346435405\\
      1.75	6.02468900888608161\\
      1.875	3.86167580075947692\\
      2	2.30961310013519583\\
      2.125	1.23293328933253443\\
      2.25	0.64350370942685831\\
      2.5	0.15915482680093759\\
      2.75	0.03651740707482166\\      
    };
    \addlegendentry{$T_{red} = 4$}

    \coordinate (lookhere) at (axis cs:2.25,8e-1);
    \spy[size=1.75cm,magnification=2.5] on (lookhere) in node at (1.1,1.1);
  \end{semilogyaxis}
\end{tikzpicture}%
	\caption{Variance of the number of trials for various \gls{snr} points for a \gls{dscf} decoder for $\mathcal{P}\left(1024,\,512\right)$ and $r=16$ with and without the proposed early-stopping mechanism.}
	\label{fig:var_exec}
\end{figure}
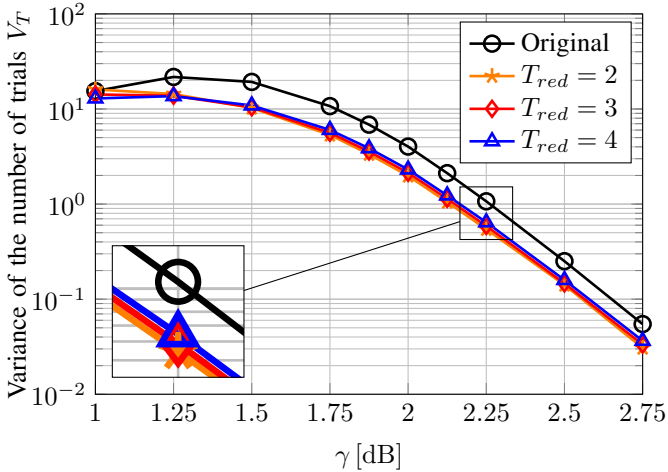

Fig.\,\ref{fig:av_exec} shows the average execution time for the proposed \gls{dscf} decoder that incorporates our early-stopping mechanism, with $T_{red}\in \{2,\,3,\,4\}$. The average execution time for the original \gls{dscf} decoding algorithm is included for comparison. Similarly, Fig.\,\ref{fig:var_exec} shows the execution-time variance for our proposed decoding algorithm and includes that of the original \gls{dscf} algorithm for comparison. In both figures, the original algorithm is depicted as a black curve with circle markers and the reminder of the curves are for our proposed modified algorithm.

From Fig.\,\ref{fig:av_exec}, it can be seen that the proposed early-stopping mechanism leads to a reduction in the average number of trials compared to the original algorithm, regardless of the reduced maximum number of trials used. This reduction is more significant at low \gls{snr} and gradually vanishes as the channel condition improves. At the \gls{snr} $\gamma=2.25$\,dB, which corresponds to a \gls{fer} of $10^{-2}$ as will be shown in Fig.\ref{fig:fer_thresh}, the reduction is of 22\% for $T_{red}=3$. Another observation is that the differences between three curves corresponding to the modified \gls{dscf} are very small and the curve representing the decoder with $T_{red}=2$ achieves the smallest average execution time.

Looking at Fig.\,\ref{fig:var_exec}, similarly to the average execution time, it can be seen that our proposed early-stopping mechanism leads to a reduction in the execution-time variance. Contrary to the reduction in the average execution time, the reduction in the execution-time variance is pretty similar across all \gls{snr} points. For the \gls{snr} point of interest, i.e., $\gamma=2.25$\,dB, the reduction is of $45\%$ for $T_{red}=3$. Again, we see that the differences between curves corresponding to $T_{red} \in \{2,\,3,\,4\}$ are very small.

\subsection{Error-Correction Performance}
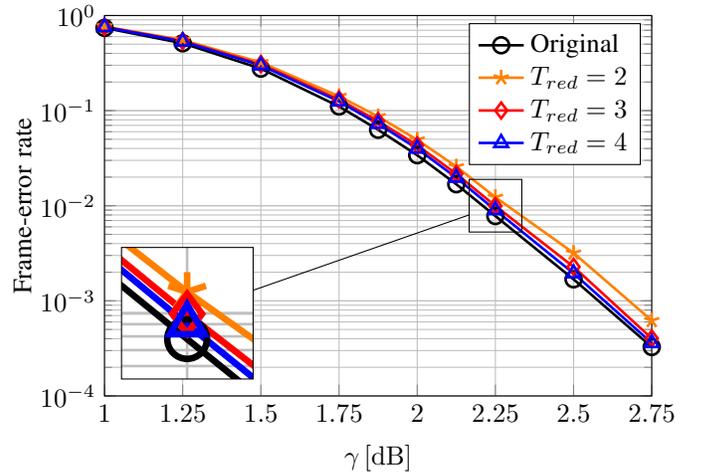
\begin{figure}[t]
  \tikzset{new spy style/.style={spy scope={%
 magnification=5,
 size=1.25cm, 
 connect spies,
 every spy on node/.style={
   rectangle,
   draw,
   },
 every spy in node/.style={
   fill=white,
   draw,
   rectangle
   }
  }
 }
} 

\begin{tikzpicture}[new spy style]

  \begin{semilogyaxis}[%
    width=\columnwidth,
    height=\plotfigureheight\columnwidth,
    xmin=1, xmax=2.75,
    xlabel={$\gamma \left[ \mathrm{dB} \right]$},
    xtick={1,1.25,...,3.0},
    ymin=1e-4, ymax=1,
    ylabel style={yshift=-0.4em},
    ylabel={Frame-error rate},
    yminorticks, xmajorgrids,
    ymajorgrids, yminorgrids,
    legend style={legend cell align=left, align=left},
    mark size=3.0pt, mark options=solid
    ]
    \addplot [color=black, line width=1.0pt, mark=o]
    table[row sep=crcr]{%
      1	0.74153\\
      1.25	0.51114\\
      1.5	0.27667\\
      1.75	0.11066\\
      1.875	0.06292\\
      2	0.03399\\
      2.125	0.01684\\
      2.25	0.007794657541720904\\
      2.5	0.001684573684937553\\
      2.75	0.000327073509771321\\
    };
    \addlegendentry{Original}

    \addplot [color=orange, line width=1.0pt, mark=star]
    table[row sep=crcr]{%
      1	0.76693\\
      1.25	0.55197\\
      1.5	0.32146\\
      1.75	0.14173\\
      1.875	0.08676\\
      2	0.0494\\
      2.125	0.02579\\
      2.25	0.0123\\
      2.5	0.003188043561427224\\
      2.75	0.000620885341104058\\ 	
    };
    \addlegendentry{$T_{red} = 2$}

    \addplot [color=red, line width=1.0pt, mark=diamond]
    table[row sep=crcr]{%
      1	0.76027\\
      1.25	0.54063\\
      1.5	0.30758\\
      1.75	0.13093\\
      1.875	0.07757\\
      2	0.04318\\
      2.125	0.02161\\
      2.25	0.009943718552990076\\
      2.5	0.002278952957853044\\
      2.75	0.000402378863843040\\ 		
    };
    \addlegendentry{$T_{red} = 3$}
    
    \addplot [color=blue, line width=1.0pt, mark=triangle]
    table[row sep=crcr]{%
      1	0.7551\\
      1.25	0.5327\\
      1.5	0.299\\
      1.75	0.12497\\
      1.875	0.07261\\
      2	0.0399\\
      2.125	0.0197\\
      2.25	0.00899\\
      2.5	0.0019465705320366580\\
      2.75	0.000359934923765783\\      	
    };
    \addlegendentry{$T_{red} = 4$}

    \coordinate (lookhere) at (axis cs:2.25,1e-2);
    \spy[size=1.75cm,magnification=2.5] on (lookhere) in node at (1.1,1.1);
  \end{semilogyaxis}
\end{tikzpicture}%
  \caption{\gls{fer} of the \gls{dscf} decoder for $\mathcal{P}\left(1024,\,512\right)$ and $r=16$ with and without the proposed early-stopping mechanism.}
  \label{fig:fer_thresh}
\end{figure}

Fig.\,\ref{fig:fer_thresh} shows the error-correction performance in terms of \gls{fer} of the modified \gls{dscf} decoder with the proposed early-stopping mechanism as well as that of the original \gls{dscf} decoder. Similarly to Figs.\,\ref{fig:av_exec} and \ref{fig:var_exec}, the original \gls{dscf} algorithm is depicted as a black curve with circle markers and the reminder of the curves are for our modified algorithm. It can be seen that a reduced maximum number of trials $T_{red}$ of 2 with the codewords identified as likely undecodable leads to a coding loss that is little under $0.1$\,dB at a \gls{fer} of $10^{-2}$ compared to the original \gls{dscf} algorithm. Increasing $T_{red}$ to 3 significantly reduces that loss. The gap between the proposed algorithm and the original one is under $0.05$\,dB at the same \gls{fer}. Increasing $T_{red}$ to 4 further reduces the error-correction loss but the improvement between $T_{red}=3$ and $T_{red}=4$ is not as important as the one created by going from 2 to 3. For reference, we note that the \gls{fer} of the \gls{dscf} decoder, with and without the proposed early-stopping algorithm, with $T=10$ falls between the \glspl{fer} of a \gls{crc}-aided \gls{scl} decoder with list sizes $L$ of 2 and 4.\medskip

Taking all three metrics into consideration---the average execution time, the execution-time variance, and the error-correction performance---, with a $\mathcal{P}\left(1024,\,512\right)$ polar code and a \gls{crc} of $r=16$ bits, using $T_{red}=3$ appears to offer the best tradeoff for a modified \gls{dscf} decoder that implements our proposed early-stopping mechanism.

\section{Conclusion}
\label{sect:conclusion}
In this work, we presented modifications to the original \gls{dscf} decoding algorithm to integrate an early-stopping mechanism that attempts to distinguish undecodable codewords from decodable. The key ingredients of this mechanism are the combination of an early-stopping metric with a pre-calculated threshold. Based on that metric, a codeword may be classified as likely undecodable, in which case the decoder attempts a reduced maximum number of trials, much lower than the initial maximal number of trials. After those trials, if decoding is not successful, the decoder stops. Compared to the original \gls{dscf} algorithm, in the region of interest for wireless communications, simulation results show that our proposed modifications could lead to reductions of $22\%$ to the average execution time and of $45\%$ to the execution-time variance at the cost of a minor error-correction loss of approximately $0.05$\,dB. By reducing the execution-time variance, using this work contributes to keeping the amount of buffering required between the modules of a receiver tractable. Future work includes adapting the proposed early-stopping mechanism to the \gls{dscf} decoding algorithm with higher orders.

\section*{Acknowledgement}
The authors would like to thank Tannaz Kalatian for helpful discussions. This work was supported by an NSERC Discovery Grant (\#651824).

\balance 
\bibliographystyle{IEEEtran}
\bibliography{IEEEabrv,ConfAbrv,refs}

\begin{thebibliography}{1}
\providecommand{\url}[1]{#1}
\csname url@samestyle\endcsname
\providecommand{\newblock}{\relax}
\providecommand{\bibinfo}[2]{#2}
\providecommand{\BIBentrySTDinterwordspacing}{\spaceskip=0pt\relax}
\providecommand{\BIBentryALTinterwordstretchfactor}{4}
\providecommand{\BIBentryALTinterwordspacing}{\spaceskip=\fontdimen2\font plus
\BIBentryALTinterwordstretchfactor\fontdimen3\font minus
  \fontdimen4\font\relax}
\providecommand{\BIBforeignlanguage}[2]{{%
\expandafter\ifx\csname l@#1\endcsname\relax
\typeout{** WARNING: IEEEtran.bst: No hyphenation pattern has been}%
\typeout{** loaded for the language `#1'. Using the pattern for}%
\typeout{** the default language instead.}%
\else
\language=\csname l@#1\endcsname
\fi
#2}}
\providecommand{\BIBdecl}{\relax}
\BIBdecl

\bibitem{arik_polariz}
E.~Ar{\i}kan, ``Channel polarization: A method for constructing
  capacity-achieving codes for symmetric binary-input memoryless channels,''
  \emph{{IEEE} Trans. Inf. Theory}, no.~7, pp. 3051--3073, Jul 2009.

\bibitem{3GPP_5G_Coding}
\BIBentryALTinterwordspacing
{3GPP}, ``{NR; Multiplexing and channel coding},'' Tech. Rep. TS 38.212, Jan
  2018, {Release 15}. [Online]. Available:
  \url{http://www.3gpp.org/DynaReport/38-series.htm}
\BIBentrySTDinterwordspacing

\bibitem{scl_5g}
F.~Ercan, C.~Condo, S.~Hashemi, and W.~Gross, ``On error-correction performance
  and implementation of polar code list decoders for {5G},'' in \emph{Ann.
  Allerton Conf. on Commun., Control, and Comput. (Allerton)}, Oct 2017, pp.
  443--449.

\bibitem{scf_intro}
O.~Afisiadis, A.~Balatsoukas-Stimming, and A.~Burg, ``A low-complexity improved
  successive cancellation decoder for polar codes,'' in \emph{Asilomar Conf. on
  Signals, Syst., and Comput. (ACSSC)}, Nov 2014, pp. 2116--2120.

\bibitem{Giard_JETCAS_2017}
P.~Giard, A.~Balatsoukas-Stimming, T.~C. M\"uller, A.~Bonetti, C.~Thibeault,
  W.~J. Gross, P.~Flatresse, and A.~Burg, ``\textsc{PolarBear}: A 28-nm
  {FD-SOI} {ASIC} for decoding of polar codes,'' \emph{{IEEE} J. Emerg. Sel.
  Topics Circuits Syst.}, vol.~7, no.~4, pp. 616--629, Dec. 2017.

\bibitem{dyn_scf}
L.~Chandesris, V.~Savin, and D.~Declercq, ``{Dynamic-SCFlip} decoding of polar
  codes,'' \emph{{IEEE} Trans. Commun.}, no.~6, pp. 2333--2345, Jun 2018.

\bibitem{simp_dscf}
F.~Ercan, T.~Tonnellier, N.~Doan, and W.~Gross, ``Simplified dynamic {SC-Flip}
  polar decoding,'' in \emph{{IEEE} Int. Conf. on Acoustics, Speech, and Signal
  Process. ({ICASSP})}, May 2020.

\bibitem{tal_constr}
I.~Tal and A.~Vardy, ``How to construct polar codes,'' \emph{{IEEE} Trans. Inf.
  Theory}, no.~10, pp. 6562--6582, Oct 2013.

\end{thebibliography}

\end{document}